\documentclass[final,5p,times,twocolumn]{elsarticle}

\usepackage{graphicx}
\usepackage{hyperref}
\journal{Preprint submitted to Nuclear Instrumentations and Methods in Physics Research, Section A}
\begin{document}

\begin{frontmatter}

\title{Evaluation of gamma-ray response of the AstroPix4 HV-CMOS active pixel sensor}

\author[huaddress]{Yusuke Suda\corref{mycorrespondingauthor}}
\cortext[mycorrespondingauthor]{Corresponding author}
\ead{ysuda@hiroshima-u.ac.jp}
\address[huaddress]{Physics Program, Graduate School of Advanced Science and Engineering, Hiroshima University, 1-3-1 Kagamiyama, Higashihiroshima, 739-8526, Hiroshima, Japan}

\author[gsfcaddress]{Regina Caputo}
\address[gsfcaddress]{NASA Goddard Space Flight Center, 8800 Greenbelt Rd, Greenbelt, 20771, MD, USA}

\author[gsfcaddress]{Daniel Violette}

\author[gsfcaddress,gwuaddress]{Grant Sommer}
\address[gwuaddress]{George Washington University, 1918 F Street, NW Washington, 20052, DC, USA}

\author[kitaddress]{Nicolas Striebig}
\address[kitaddress]{ASIC and Detector Laboratory, Karlsruhe Institute of Technology, Hermann-von-Helmholtz-Platz 1, Karlsruhe, D-76344, Baden-Württemberg, Germany}

\author[anladdress]{Manoj Jadhav}
\address[anladdress]{Argonne National Laboratory, 9700 S. Cass Avenue, Lemont, 60439, IL, USA}

\author[huaddress]{Yasushi Fukazawa}

\author[gsfcaddress]{Carolyn Kierans}

\author[kitaddress]{Richard Leys}

\author[anladdress]{Jessica Metcalfe}

\author[huaddress]{Norito Nakano}

\author[kitaddress]{Ivan Peri\'{c}}

\author[gsfcaddress]{Jeremy S. Perkins}

\author[iseeaddress,kmiaddress]{Hiroyasu Tajima}
\address[iseeaddress]{Institute for Space–Earth Environmental Research, Nagoya University, Furo-cho, Chikusa-ku, Nagoya, 464-8601, Aichi, Japan}
\address[kmiaddress]{Kobayashi-Maskawa Institute for the Origin of Particles and the Universe, Nagoya University, Furo-cho, Chikusa-ku, Nagoya, 464-8602, Aichi, Japan}

\begin{abstract}
AstroPix is a novel high-voltage CMOS active pixel sensor being developed for a next generation gamma-ray space telescope, AMEGO-X.
To meet AMEGO-X instrument requirements, AstroPix must achieve full depletion of its $500~\rm{\mu m}$ thick, $500~\rm{\mu m}$-pitch pixels.
It must be sensitive to gamma rays in the range of $25-700$~keV, with the energy resolution at 122~keV of $<10$~\%.
Furthermore, given the space-based nature of AMEGO-X, the power consumption of AstroPix needs to be lower than $1.5~\rm{mW/{cm}^2}$.
We report the gamma-ray response of the latest version of AstroPix, AstroPix4.
The chip contains $16\times 13$ array of $500~\rm{\mu m}$-pitch pixels.
The power consumption is estimated to be about $2~\rm{mW/{cm}^2}$, which is approximately half the power of the previous AstroPix version.
The input capacitance is reduced, allowing for the detection of the 14~keV photopeak from $\rm{^{57}Co}$ and a moderate energy resolution of 14\% at 122~keV.
The dynamic range is estimated to be in the range from 14~keV to $\sim250$~keV.
We found that the sensor depletion layer expands as expected and the measured depletion depth is approximately $90~\rm{\mu m}$ when biased at $-240$~V.
\end{abstract}


\end{frontmatter}


\section{Introduction}
High energy astronomical objects such as gamma-ray bursts, active galactic nuclei, and others contain rich physics which cannot be studied on Earth.
Those objects typically emit MeV gamma-rays which often dominate other wavelengths of electromagnetic radiation, and thus MeV gamma-ray observations are important.
The rise of multi-messenger astronomy has further increased the demand for MeV gamma-ray observations.
Therefore, the All-sky Medium Energy Gamma-ray Observatory eXplorer (AMEGO-X) is proposed as a future MeV mission concept \cite{caputo}.
Its gamma-ray telescope has four silicon tracker towers which require a huge silicon area of $24~\rm{m^2}$ in total.
Due to the fact that Compton scattering plays the main role in MeV photon-detector interactions, the silicon sensor must be thick and fully depleted with a low noise floor and moderate energy resolution.
Since the available power is limited on satellites, low power consumption is required for the silicon sensor.
The requirements for AMEGO-X are summarized in Table~\ref{tab:req} based on \cite{brewer}.
\begin{table}[htbp]
  \caption{Requirements for AstroPix. The energy resolution value is expressed in units of full width at half maximum (FWHM).}
  \label{tab:req}
  \centering
  \begin{tabular}{lc}
    \hline
    Pixel pitch & $500\times500~\rm{\mu m^2}$\\
    Thickness & $500~\rm{\mu m}$\\
    Dynamic range & 25~keV - 700~keV\\
    Energy resolution & $<10\%$ (FWHM) at 122 keV\\
    Power consumption & $<1.5~\rm{mW/{cm}^2}$\\
    \hline
  \end{tabular}
\end{table}

Toward AMEGO-X, we have been developing a novel high-voltage CMOS (HV-CMOS) active pixel sensor, AstroPix.
In this paper, we explain the AstroPix sensor and its latest version, AstroPix4, in Section \ref{sec:astro}, and present the energy calibration and depletion depth measurements in Sections \ref{sec:calib} and \ref{sec:depth}, respectively.
We summarize the obtained performance of AstroPix4 and conclude with future prospects in Section \ref{sec:conc}.

\section{AstroPix}
\label{sec:astro}
The AstroPix detector design has leveraged experience gained through the developments of both MuPix and ATLASPix \cite{peric, pericberger}.
The pixel CMOS circuitry is implemented inside a deep N-well on P-substrate.
The sensor aims to achieve full depletion by applying an HV bias via a guard-ring.
Electrons produced by gamma rays or charged particles in the depletion layer drift toward the deep N-well.
Collected signal goes through an intra-pixel charge-sensitive amplifier (CSA) and comparator, then a Time-over-Threshold (ToT) measurement is performed in the digital periphery located at the bottom of the chip.
A charge injector is implemented in each pixel, allowing to produce a test pulse and the injected charge is adjusted via the injection voltage.

We developed and studied three versions of AstroPix; AstroPix1, AstroPix2, and AstroPix3 \cite{steinhebel, suda, v3}.
The pixel pitch gradually increased and reached the target value of $500~\rm{\mu m}$ in AstroPix3.
The energy resolution at 122~keV was not yet measured, but in AstroPix3 at 60~keV it was measured to be 6.2~keV (FWHM) with 44.4\% of pixels satisfying the 10\% resolution requirement.
Larger pixels and improvements in the chip design have lead to a gradual reduction in power consumption, reaching $4.12~\rm{mW/{cm}^2}$ in AstroPix3.

\begin{figure}[htb]
\begin{center}
\includegraphics[width=0.6\linewidth]{
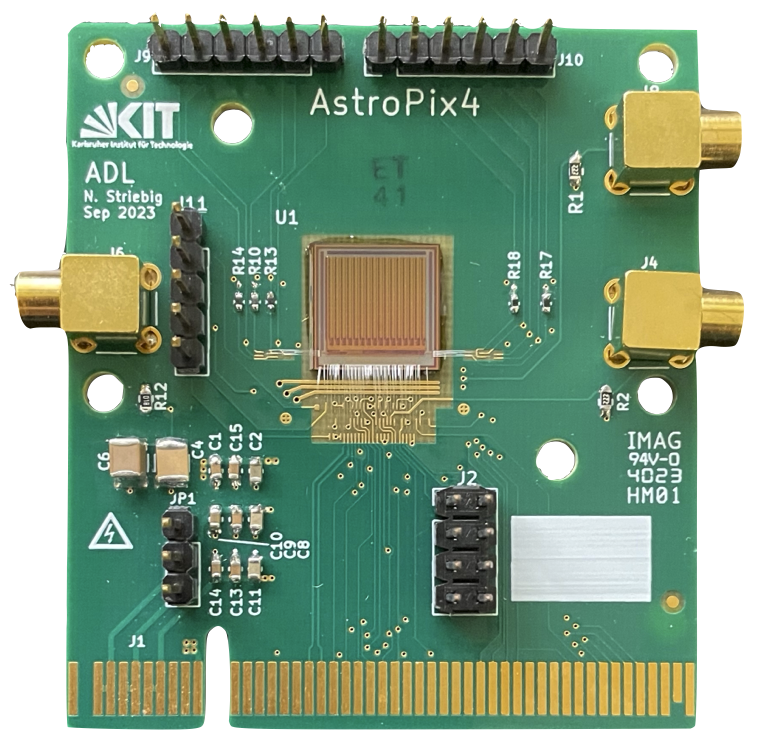}
\caption{AstroPix4 chip and its carrier board. Designed by Karlsruhe Institute of Technology in Germany.}
\label{fig:v4}
\end{center}
\end{figure}

In the fourth version of AstroPix, AstroPix4, the chip is $725~\rm{\mu m}$ thick and contains an array of $16\times 13$, $500~\rm{\mu m}$-pitch pixels (Fig.~\ref{fig:v4}).
It was fabricated on a $(200 - 400)~\rm{\Omega\cdot cm}$ wafer using TSI Semiconductors' 180~nm process.
There are four notable design changes from AstroPix3 (details can be found in \cite{striebig}): 1. Reduced input capacitance. 2. Pixel-by-pixel comparator threshold tuning functionality. 3. Individual hit buffer implementation. 4. Improved time stamp structure.
The first one was achieved by optimizing the routing and minimizing the metal-to-N-well capacitance, leading to a lower noise floor and better energy resolution.
The second item is to suppress ToT spread among pixels, but was not used in this work.
In the previous AstroPix versions, the comparator output of each hit in one row and column are OR'd, which resulted in hit misidentification if there are multiple hits in one row or column.
The third improvement can solve this hit misidentification problem since each pixel has its own hit buffer.
The last change aimed to reduce the power consumption by employing a 20~MHz clock with 4 bit flash Time-to-Digital Converters instead of using a 200~MHz clock which was used in up to AstroPix3.
The power consumption of AstroPix4 is about $2~\rm{mW/{cm}^2}$, approximately half the power of AstroPix3.

\section{Energy calibration}
\label{sec:calib}
\begin{figure}[htb]
\begin{center}
\includegraphics[width=0.9\linewidth]{
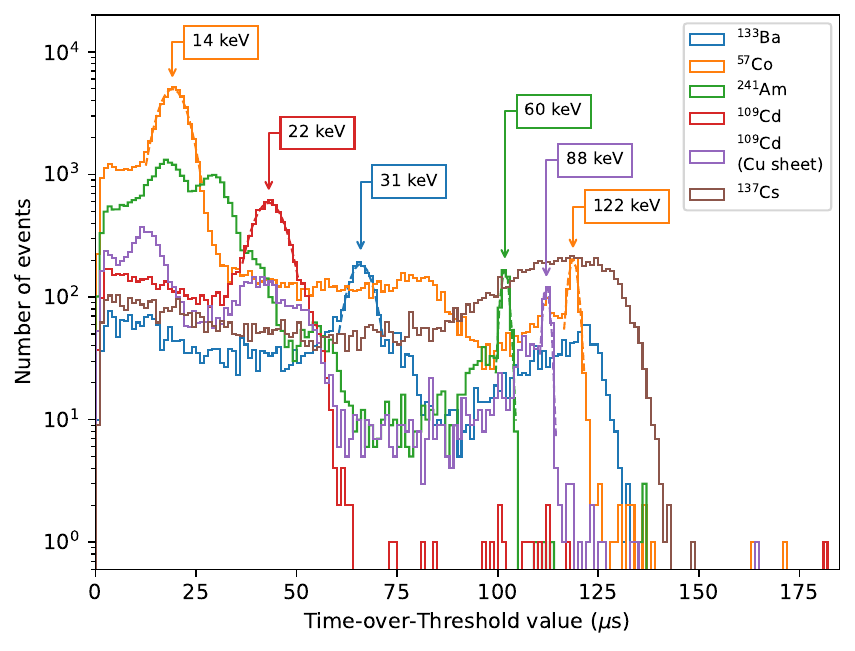}
\caption{Energy spectra of five radioisotopes obtained from a single pixel. The purple spectrum was taken with a copper sheet placed between $\rm{^{109}Cd}$ and the chip to suppress the hit rate due to lower energy photons. Dashed lines represent fitted Gaussian functions.}
\label{fig:totspec}
\end{center}
\end{figure}

To evaluate the gamma-ray response, we conducted ToT measurements using various radioisotopes.
The operating bias voltage in this Section was -200~V supplied by a KEITHLEY 2450 SourceMeter with the current draw of $\sim50$~nA.
Fig.~\ref{fig:totspec} shows the energy spectra of five radioisotopes ($\rm{^{133}Ba}$, $\rm{^{57}Co}$, $\rm{^{241}Am}$, $\rm{^{109}Cd}$, and $\rm{^{137}Cs}$) in units of ToT obtained from a single pixel.
Six emission lines (14~keV, 22~keV, 31~keV, 60~keV, 88~keV, and 122~keV) are clearly visible.
Due to the reduced input capacitance, the photopeak of 14~keV which was below the noise floor of AstroPix3 can be found easily.
A cut-off feature at around $130~\rm{\mu s}$ in $\rm{^{133}Ba}$ can be seen.
As mentioned later, this feature is caused by both 303~keV and 356~keV photons from $\rm{^{133}Ba}$, which create Compton edges at 164~keV and 207~keV, respectively.
On the other hand, the cut-off feature at around $140~\rm{\mu s}$ in $\rm{^{137}Cs}$ is due to the saturation of the CSA output which is expected to happen at around 250~keV.

\begin{figure}[htb]
\begin{center}
\includegraphics[width=1\linewidth]{
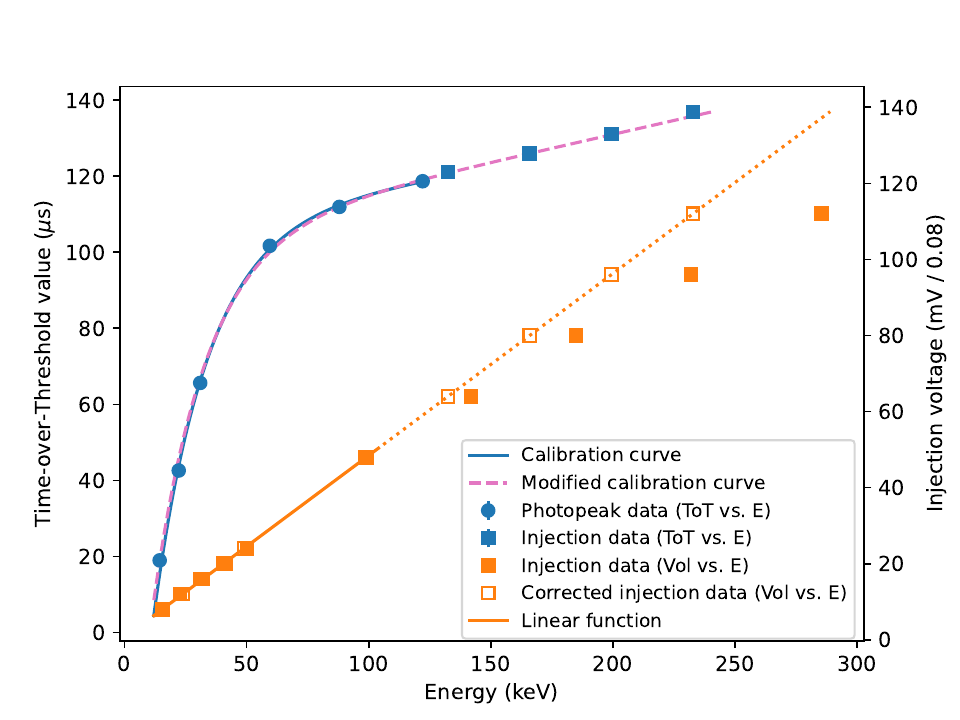}
\caption{Energy calibration curve (blue) and the relationship between the injection voltages and their calibrated energies (orange) for a single pixel.}
\label{fig:calib}
\end{center}
\end{figure}

\begin{figure}[htb]
\begin{center}
\includegraphics[width=1\linewidth]{
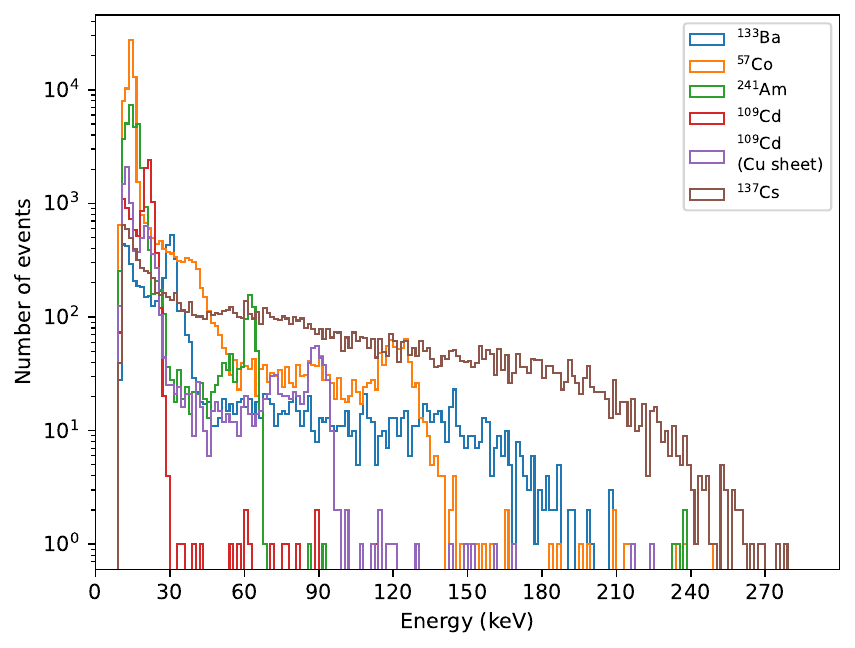}
    \begin{picture}(0,0)
        \put(-45,110){\includegraphics[width=0.22\textwidth]{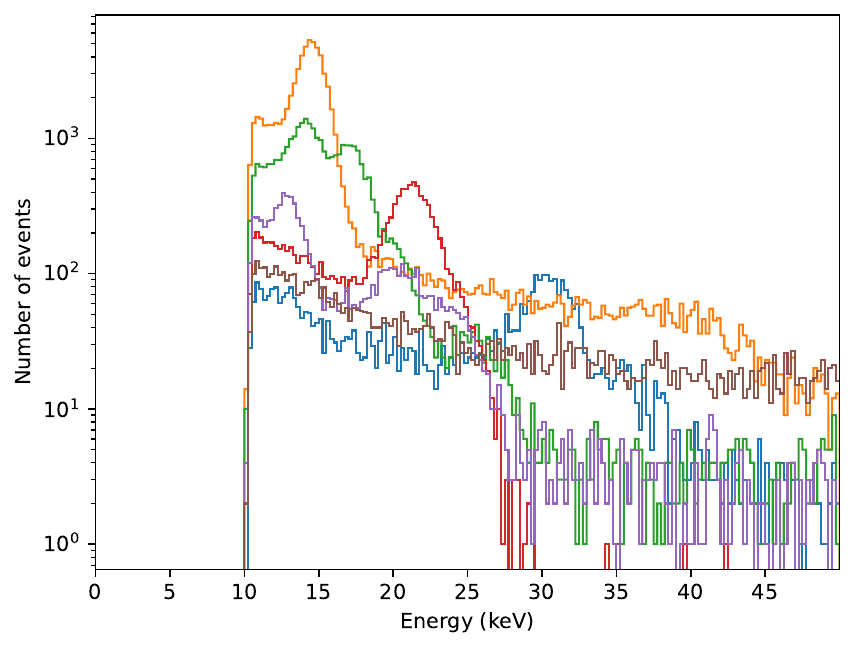}}
    \end{picture}
\caption{Energy spectra obtained from a single pixel. The inset illustrates a magnified view of the lowest energy region.}
\label{fig:enespec}
\end{center}
\end{figure}

\begin{figure}[htb]
\begin{center}
\includegraphics[width=0.8\linewidth]{
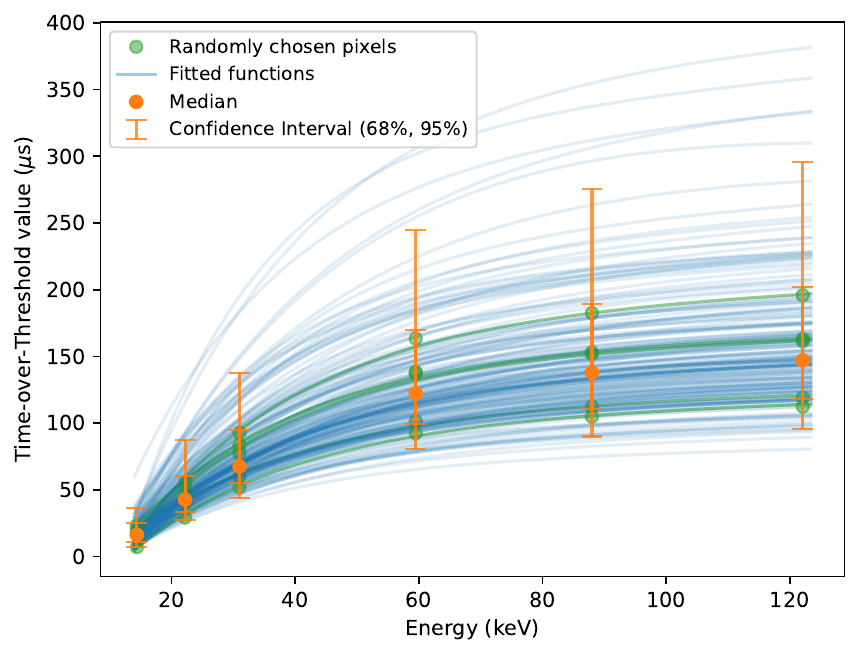}
\caption{Energy calibration curves for all calibrated pixels. Green points and curves show the Gaussian fitted mean ToT values of randomly chosen pixels and their fitted calibration curves, respectively. Blue curves represent calibration curves for the other pixels. Orange points and error bars are medians of peak ToT distribution and their confidence intervals (68\% and 95\%).}
\label{fig:calibcurves}
\end{center}
\end{figure}

We performed an energy calibration on pixels which have six photopeaks.
The fraction of calibrated pixels is 88.3\% (181 pixels, excluding 3 masked pixels).
The calibration curve for the same pixel used in Fig.~\ref{fig:totspec} is shown in Fig.~\ref{fig:calib}.
We employ an empirical function ($ToT = aE+b[1-\exp(-E/c)]+d$, where $E$ is the photopeak energy) to fit the data points, shown in blue.
To investigate the high energy end of the dynamic range, we would require higher energy photopeaks ($>122$~keV) or beam testing to extend the calibration curves up to the CSA saturation energy.
Fig.~\ref{fig:calib} also displays the relationship between the injection voltages and their calibrated energies.
A linear behavior is observed up to around 100~keV and then the data points deviate afterwards as the calibration curve has some uncertainty above 122~keV which is the highest photopeak data point.
However, we can correct those injection data above 122~keV under the assumption of the linearity between injection voltage and energy ($Q = CV \propto E$), shown as blank squares.
By adding those corrected injection data points, the calibration curve can be modified and is indicated by the pink dashed line.
Fig.~\ref{fig:enespec} represents calibrated energy spectra for the ones in Fig.~\ref{fig:totspec}, obtained from the modified calibration curve.
All photopeaks are well reconstructed and the above mentioned Compton edge formed by both 303~keV and 356~keV photons can be confirmed in $\rm{^{133}Ba}$.
The drop in the highest energy regime in $\rm{^{137}Cs}$ shows that the CSA indeed saturates at around 250~keV and thus we confirm the dynamic range ranging from 14~keV to $\sim250$~keV.
Fig.~\ref{fig:calibcurves} shows all calibration curves for the calibrated pixels.
The median and 68\% confidence interval of the parameters in the empirical function are $a=0.12^{+0.06}_{-0.04}$, $b=207^{+81}_{-43}$, $c=28^{+4}_{-3}$, and $d=-64^{+13}_{-37}$.

\begin{figure}[htb]
\begin{center}
\includegraphics[width=1\linewidth]{
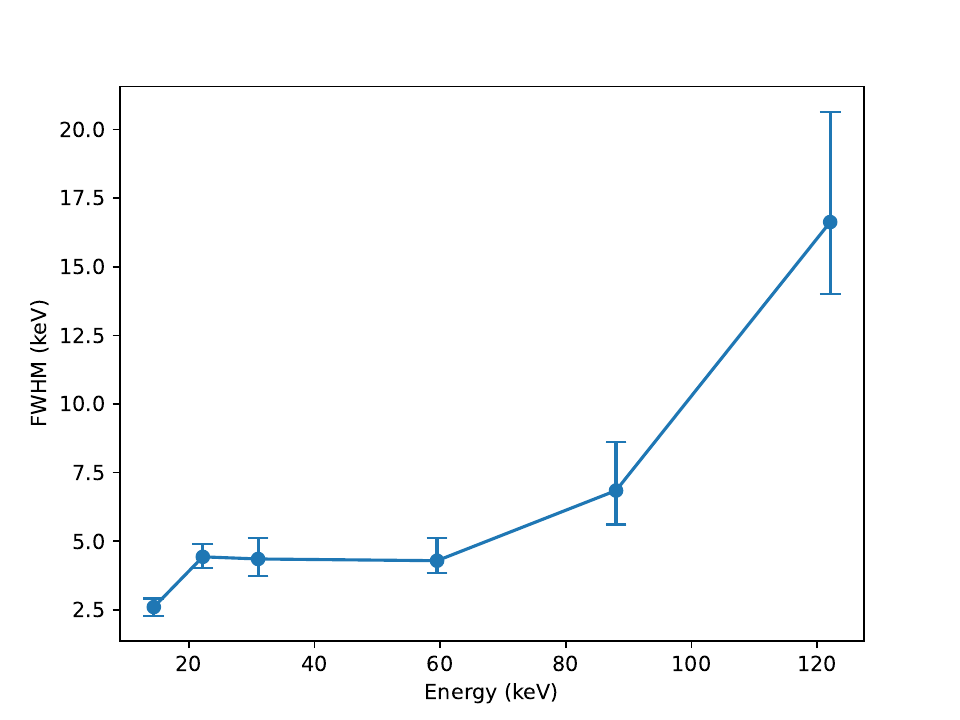}
    \begin{picture}(0,0)
        \put(-93,82){\includegraphics[width=0.24\textwidth]{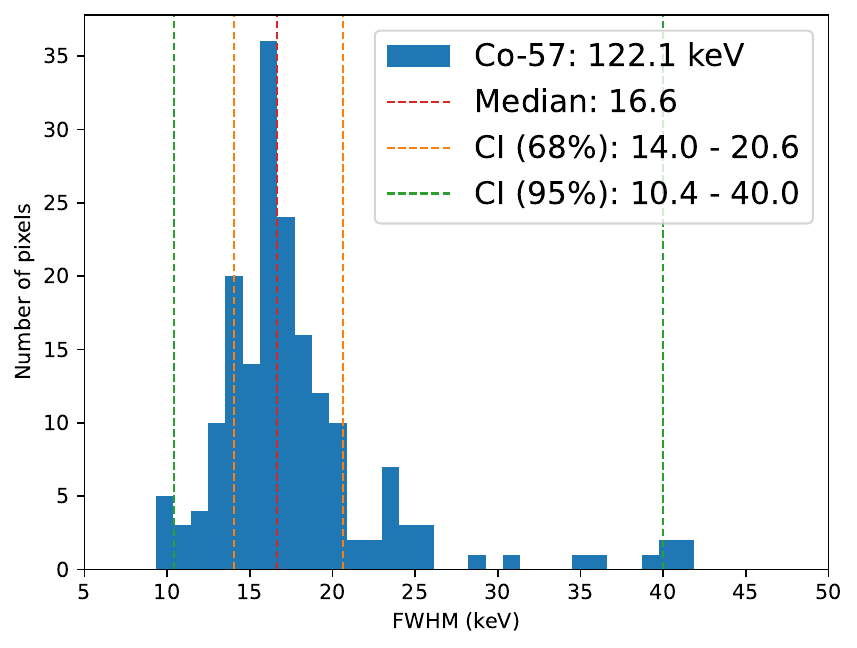}}
    \end{picture}
\caption{Energy dependence of energy resolutions (FWHM). The inset displays the energy resolution distribution of 122~keV photons. CI stands for confidence interval.}
\label{fig:ereso}
\end{center}
\end{figure}

As shown in the inset of Fig.~\ref{fig:ereso}, the energy resolution at 122~keV can be estimated to be $16.6^{+4.0}_{-2.6}$~keV.
This is larger than those at lower energies as observed in Fig.~\ref{fig:ereso}.
The reason is not identified, but it could be that the achieved depletion depth described in Section~\ref{sec:depth} is not thick enough to fully absorb the deposited energy as 122~keV photoelectrons have a $100~\rm{\mu m}$-long path length in silicon.
The energy resolution at 60~keV was measured as $4.3^{+0.8}_{-0.4}$~keV, which is a 30\% improvement compared to that in AstroPix3 \cite{suda}.

\section{Depletion depth measurements}
\label{sec:depth}
Even though the resistivity of the chip in this work is not high enough to allow full depletion (a $(5-20)~\rm{k\Omega\cdot cm}$ wafer is required), it is important to estimate the depletion depth in order to deepen our understanding of the sensor.
We measured the depletion depth following the analysis procedure described in Section 4 of \cite{suda}, namely, by measuring the ToT spectrum of $\rm{^{241}Am}$, counting the number of 60~keV photons, and deriving the depletion depth from the photopeak event rate.
The measurements were conducted with a copper sheet inserted between the chip and the $\rm{^{241}Am}$ source to filter out lower energy photons.
The measured ToT spectrum was fitted with a combined function of three components (primary Gaussian for the photopeak, secondary Gaussian for the Compton back-scattered peak around 50~keV, and error function for the lower energy flat spectral feature).
By subtracting the two extra components from the spectra, a Gaussian-like photopeak distribution was obtained, and the number of the photopeak events was calculated by integrating the subtracted spectrum.
While calculating the rate, we considered an effective time to account for the dead time in the readout (the correction factor to the observed time is around 0.91) and an effective volume to address the non-negligible path length of 60~keV photoelectrons (the correction factor to the nominal volume is ranging from 0.83 to 0.87 depending on the bias voltage, based on Geant4-based simulations).

\begin{figure}[htb]
\begin{center}
\includegraphics[width=0.8\linewidth]{
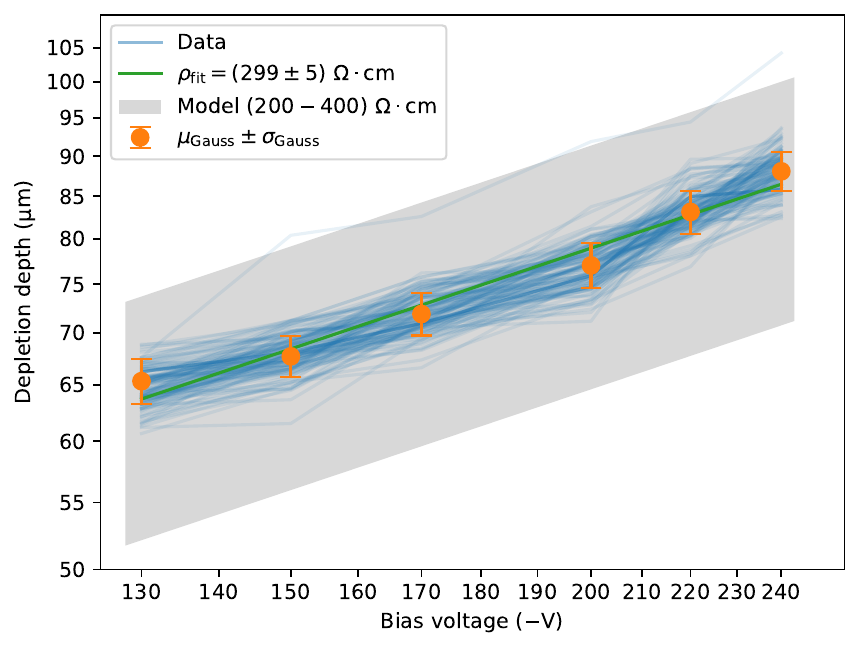}
\caption{Depletion depths as a function of bias voltage. The lines in blue show the depletion depths for each pixel. The orange points represent the Gaussian means of the depletion depth distribution and the error bars are the fitted sigmas ($65\pm2, 68\pm2, 72\pm2, 77\pm2, 83\pm3, 88\pm2~\rm{\mu m}$ at 130, 150, 170, 200, 220, 240 V, respectively). The green line is the fitted PN junction model. The gray band displays the model band with the uncertainty in the resistivity.}
\label{fig:depths}
\end{center}
\end{figure}

The bias voltage dependence of the measured depletion depths is shown in Fig.~\ref{fig:depths}.
We compare the data points with a simple PN junction model 
        $d = \sqrt{2\epsilon\mu\rho\left(V_{\rm{bias}} + V_{\rm{built\_in}} \right)}$,
where $d$ is the theoretical depletion depth, $\epsilon$ the permittivity, $\mu$ the hole mobility, $\rho$ resistivity, $V_{\rm{bias}}$ the bias voltage, and $V_{\rm{built\_in}}$ the built-in potential.
The data points are well within the model band with the uncertainty in the resistivity, and follow the overall tendency of the model.
From a fit using each representative data point in orange, we obtained the estimated substrate resistivity of $(299\pm5)~\rm{\Omega\cdot cm}$.

\section{Conclusions and prospects}
\label{sec:conc}
We have been developing AstroPix toward the next generation MeV gamma-ray satellite, AMEGO-X.
In this paper, we presented the gamma-ray response of its latest version, AstroPix4.
The decrease in the input capacitance led to a lower noise floor, allowing us to detect the 14~keV photopeak, which was not visible in the previous version.
We introduced a new method to modify the energy calibration curve by utilizing injection data, which allows to explore higher energy regime where virtually no photopeak is available due to low cross section.
With this method, the dynamic range of AstroPix4 was estimated to be in the range from 14~keV to $\sim250$~keV.
To achieve the required dynamic range, the next version of AstroPix, AstroPix5, will be equipped with two test columns with high dynamic range charge-sensitive amplifier that should not saturate until $>700$~keV.
The energy resolution at 60~keV gets improved by 30\% compared to AstroPix3, but that at 122~keV is 14\% which has not yet met the requirement.
This could be due to the inefficiency of detecting 122~keV photoelectrons which have a corresponding path length of $100~\rm{\mu m}$.
Further reduction in the input capacitance by reducing metal layers in AstroPix5 would allow us to achieve the requirement in energy resolution.

Since the chip was fabricated in a medium-resistivity wafer, the chip in this work has not yet reached full depletion.
The measured depletion depth is approximately $90~\rm{\mu m}$.
AstroPix5 will be implementing a new guard ring design proposed in \cite{zhang} which is expected to result in a higher breakdown voltage.
We plan to fabricate AstroPix5 in higher resistivity wafers, anticipating that it can achieve a much larger depletion depth.

\section*{Acknowledgments}
\label{sec:ack}
The authors gratefully acknowledge financial support from NASA (18-APRA18-0084).
This work was supported by JSPS KAKENHI Grant Numbers JP23K13127, JP23H04897, JP25K01028.

\end{document}